\documentclass{eptcs}
\providecommand{\event}{TTC 2011}

\usepackage{oz2}
\usepackage{alltt}
\usepackage{graphicx}

\title{Saying Hello World with UML-RSDS -- A Solution to the 2011 Instructive Case}
\author{K. Lano, S. Kolahdouz-Rahimi\\
{Dept. of Informatics, King's College London, Strand, London, UK\thanks{Research supported by the HoRTMoDA EPSRC project}}
\email{kevin.lano@kcl.ac.uk}}

\def\titlerunning{Hello World Case in UML-RSDS}
\def\authorrunning{K. Lano, S. Kolahdouz-Rahimi}

\begin{document}
\maketitle

\begin{abstract}
In this paper we apply the UML-RSDS notation
and tools to the ``Hello World" 
case studies and explain the underlying development
process for
this model transformation approach.
\end{abstract}

\section{Specification of model transformations}

In UML-RSDS a transformation specification is written in
first-order logic and OCL, and 
consists of the following predicates:
\begin{enumerate}
\item A global specification, $Cons$, of a model
transformation, expresses in a platform-independent
manner the overall effect of the transformation,
as a relation between the source and target models.
It is intended to
hold true at termination of the transformation.
\item A predicate $Asm$ expresses the assumptions
made about the source and target models at the start
of the transformation, for example, that the target
model is empty and that the source model is syntactically
correct wrt the source language. 
\end{enumerate}
The specification is therefore independent of any
specific model transformation implementation 
language, and can be used as the basis for development
in many such languages. By making explicit the semantic
assumptions on source and target models, the 
specification assists in the verification (formal or 
informal) of model transformations. 

$Cons$ can often be
written in {\em conjunctive-implicative form} \cite{mtpat}, 
as a 
conjunction of constraints of the form
\[ \forall s : S \cdot SCond ~~implies~~ \exists t : T \cdot Post \]
where $S$ is a source language entity and $T$ is a 
target language entity. This pattern is applicable to
re-expression transformations such as model
migrations, and to abstraction and refinement 
transformations. 

The patterns assist in the derivation of explicit 
PSM designs from the specification, consisting of 
a sequence of {\em phases}, which apply specific
rules or operations to achieve the specification
constraints. Provided that
the updates defined in $Post$ do not affect the
data read in $SCond$ or $Post$, and that 
the extent of $S$ is fixed throughout the transformation
$(*)$,
then a constraint of the above form can be
implemented by an iteration
\begin{alltt}
for \(s : S\) do \(s.op()\)
\end{alltt}
where $op$ implements the constraint for a 
particular $S$ object.

This iteration constitutes a single phase in the design.
The possible orderings of phases are determined by
defining a partial order over the target language
entities: $T_1 < T_2$ if $T_1$ is used in $Cons$ to define
a feature of $T_2$ (or a feature
of a subclass of $T_2$). Any 
phase that creates $T_2$ instances must therefore be
preceded by all phases that create $T_1$ instances.

The restriction $(*)$
 is termed the {\em non-interference} condition.

The iterative phase activities derived from the
constraints are also
terminating and they establish the truth of their
corresponding constraint, by construction. 
The PSM design is derived from the constraints,
together with an executable Java implementation, 
using the UML-RSDS toolset \cite{Lano10ifm}.
The resulting executable is a stand-alone implementation
of the transformation, operating upon simple text format
files defining input and output models.

\section{Simple transformation tasks}

Here we give the specifications and implementations of
the simple transformation tasks in \cite{helloworldcase}. 
All of these tasks satisfy the restrictions described 
above, so they can be specified and designed directly 
in UML-RSDS.

\subsection{Hello world transformation}

This has the global specification ($Cons$ predicate):
\[ \exists g : Greeting \cdot g.text = ``Hello" ~~and~ \\
\t2 \exists p : Person \cdot  g.whom = p ~~and~~ p.name = ``World" \]

This predicate is coded in UML-RSDS as
the only postcondition
\begin{alltt}
  \(Greeting{\fun}exists( g | g.text = ``Hello"  \& \)
  \(    Person{\fun}exists( p | g.whom = p  \&  p.name = ``World" ) )\)
\end{alltt}
 of a use case which represents the transformation. From this an implementation is automatically generated in
Java.

\subsection{Graph properties}

Figure \ref{graphqmm} shows the basic graph
metamodel in the UML-RSDS tools, and the generated
design and Java code of the specification.

We assume that the following constraint $Asm0$
of the source model holds:
\[ \forall g : Graph \cdot~ g.edges.src ~\subseteq~ g.nodes ~~and~~ g.edges.trg ~\subseteq~ g.nodes \]

The queries are simple examples of abstraction
transformations, and can be specified as follows:

The constraint
\begin{alltt}
  \(IntResult{\fun}exists( r | r.num = nodes{\fun}size() ) \)
\end{alltt}
on $Graph$ expresses that for each graph 
there is a result object
recording the number of nodes in the graph. An 
operation $op1()$ is generated to implement the
constraint.

Likewise for the other queries:
\begin{alltt}
  \(IntResult{\fun}exists( r | r.num = edges{\fun}select(src = trg \& trg \neq \{\}){\fun}size()  ) \)
\end{alltt}
counts the number of looping edges in each graph, 
and is
implemented by an operation $op2()$.

\begin{alltt}
  \(IntResult{\fun}exists( r | r.num = g.edges{\fun}select(src = \{\}  or  trg = \{\}){\fun}size() ) \)
\end{alltt}
counts the number of dangling edges and is
implemented by an iteration of an operation $op3()$
on graphs.

\begin{alltt}
  \(IntResult{\fun}exists( r | r.num = (g.nodes - (g.edges.src \cup g.edges.trg)){\fun}size() ) \)
\end{alltt}
counts the number of nodes that are not the source
or target of any edge. $-$ denotes set subtraction and
$\cup$ set union. This is implemented by an
operation $op4()$.

\begin{figure}[htbp]
\centering
\includegraphics[width=3.2in]{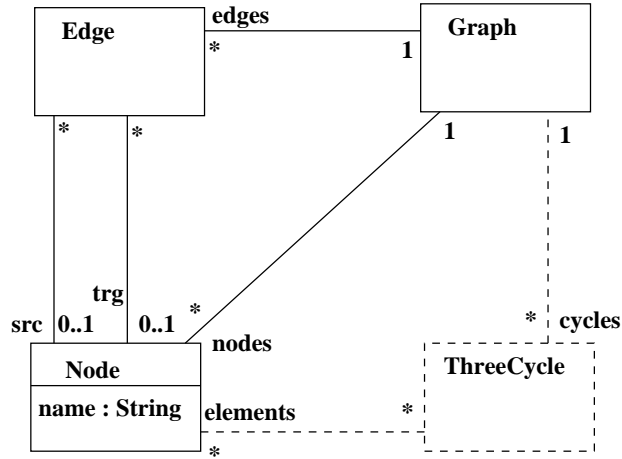}
\caption{Extended graph metamodel}
\label{3cyclesmm}
\end{figure}

We extend the final query problem by defining an
auxiliary entity which records the 3-cycles in the 
graph (Figure \ref{3cyclesmm}).

The specification $Cons$ of this transformation
then defines how unique elements of
$ThreeCycle$ are derived from the graph, and 
returns the cardinality of this type in the end state
of the transformation:
\[ (C1):~ \\
\t1 e1 : edges ~\&~ e2 : edges ~\&~ e3 : edges ~\&\\
\t1 e1.trg = e2.src ~\&~ e2.trg = e3.src~ \&~ e3.trg = e1.src ~\&\\
\t1 (e1.src \cup e2.src \cup e3.src){\fun}size() = 3 ~~\implies \\
\t2   ThreeCycle{\fun}exists1( tc | tc.elements = (e1.src \cup e2.src \cup e3.src) ~\&~ tc : cycles ) \]

\[ (C2):~  IntResult{\fun}exists( r | r.num = cycles{\fun}size() ) \]
Both constraints are on $Graph$.

The order of nodes in a cycle is not distinguished by
$C1$, if this was required then $elements$ should be
ordered (a sequence). Because of $Asm0$, each three-cycle
will consist of nodes in a single graph.
The unique existential quantifier $\exists_1$ specifies
that there must exist exactly one object satisfying the
quantified properties, ie, duplicated cycles are not
included in $cycles$.

Each constraint is refined by a specific phase in the
design. 
The $exists1$ quantifier is implemented by checking that
there is no existing $ThreeCycle$ with the required
property, before creating such an element.


An alternative approach would be to evaluate the set of
three cycles in a single expression:
\[ edges{\fun}collect( e1, e2, e3 | \{ e1, e2, e3 \} ){\fun}asSet(){\fun}select( s | \\
\t4 s{\fun}size() = 3 ~\&~ s.src = s.trg ){\fun}size() \]
but we consider that the approach using $ThreeCycle$
is more clear.

\subsection{Reverse edges}

The global specification
$Cons$ for this transformation is:
\[ src = trg{\mbox{@}}pre ~\&~ trg = src{\mbox{@}}pre \]
on $Edge$.
The suffix ${\mbox{@}}pre$ denotes the value of the
expression at the start of the transformation. This is
the usual style of specification for update-in-place
transformations.




\subsection{Simple migration}

The metamodels for this re-expression transformation
are shown in Figure 
\ref{smigmm}, together with extracts from
example input and 
output models (on the left and right hand sides,
respectively).

We make the additional assumption $Asm1$ that the
target model is empty at the start of the transformation:
\[ ModelElement2 = \{\} \]

We can specify this transformation by three constraints,
defined as the postconditions of a single use
case of the system:
\[ (C1):~ Node2{\fun}exists( n2 | n2.id2 = id1 ~\&~ n2.text = name ) \]

\[ (C2):~ Edge2{\fun}exists( e2 | e2.id2 = id1 ~\&~ e2.text = ``" ~\&\\
\t2 e2.src2 = Node2[src1.id1] ~and~ e2.trg2 = Node2[trg1.id1] ) \]
$C1$ is a constraint on $Node1$, and $C2$ on $Edge1$.
$Node2[src1.id1]$ denotes the set of $Node2$ objects
with primary key $id2$ value in the set $src1.id1$.

\[ (C3):~  Graph2{\fun}exists( g2 | g2.id2 = id1 ~\&\\
\t2 g2.gcs = Node2[nodes.id1] ~\cup~ Edge2[edges.id1] ) \]

$C3$ is a constraint on $Graph1$.
A design can be automatically generated from these
constraints, this implements each constraint by a separate
phase in a three-phase algorithm.
The ordering of the phases follows from the ordering of
the entities $Node2 < Edge2 < Graph2$ in the target
language, based upon the dependencies between these
entities in the specification constraints ($Edge2$ 
instances depend upon $Node2$ instances, etc).

\subsection{Delete nodes}

The global specification of this update-in-place
transformation can be written as:
\[ edges{\fun}select(src.name = n1~ or~ trg.name = n1){\fun}isDeleted() ~\& \\
nodes{\fun}select(name = n1){\fun}isDeleted() \]
on $Graph$.
The predicate also serves as the definition of an
operation $remove(s : String)$
of $Graph$ that implements the transformation.
Since edges depend on nodes, edges are deleted before
nodes (the reverse to the ordering used in construction
of a model). 

\subsection{Insert transitive edges}

This can be considered as a simple example of a
quality-improvement model transformation. Such
transformations are typically update-in-place
transformations, and have an associated quality measure
$Q : \nat$ on the models, used to show termination of
the transformation. The transformation aims 
to reduce $Q$ to 0 in the target
model. In this case $Q$ is the number of pairs of distinct
non-dangling edges $e1$, $e2$ of the source model
with $e1.trg = e2.src$ and with no 
existing edge from $e1.src$ to $e2.trg$.

Under the assumption $Asm2$ that there are not already
any duplicate edges in the graph:
\[ \forall e1, e2 : Edge \cdot e1.src = e2.src ~implies~ e1.trg \neq e2.trg \]
the specification of this transformation can be written as:
\[ (Cons): ~\\
\t1  e1 : edges{\mbox{@}}pre ~\&~ e2 : edges{\mbox{@}}pre ~\& \\
\t1 e1.trg = e2.src ~\&~ e1.src \neq \{\}  ~\&\\
\t1 e1.trg \neq \{\} ~\&~ e2.trg \neq \{\} ~\implies\\
\t3 Edge{\fun}exists1( e3 | e3.src = e1.src ~\&~ e3.trg = e2.trg ~\&~ e3 : edges ) \]
on $Graph$.
This satisfies the non-interference condition (since the
created $e3$ edges are distinct and are not included in
the sets of edges being iterated over), so
permitting an implementation using fixed iterations.
If instead the {\em transitive closure} $R^+$ of $R$ was
required, $Cons$ would use $edges$ instead of
$edges$@$pre$, and a more complex implementation
strategy would be required, using repeated
iteration until a fixed point is reached \cite{mtpat}.


\section{Conclusion}

We have shown that UML-RSDS can specify the
case study transformations in a direct manner
as high-level specifications, from which designs and
executable implementations can be automatically
generated.
UML-RSDS has the advantage of using standard UML
and OCL notations to specify transformations, reducing
the cost of learning a special-purpose transformation
language. Our method also has the advantage of making
explicit all assumptions on models (eg, $Asm0$ above)
and providing global specifications ($Cons$, $Asm$) 
of transformations, independent of specific
rules.

Further work includes linking UML-RSDS to Eclipse/EMF
to enable the use of ecore metamodels and import/export
of Eclipse/EMF models.

\appendix

\section*{Appendix A: Transforming specific models}

\begin{figure}[htbp]
\centering
\includegraphics[width=4.1in]{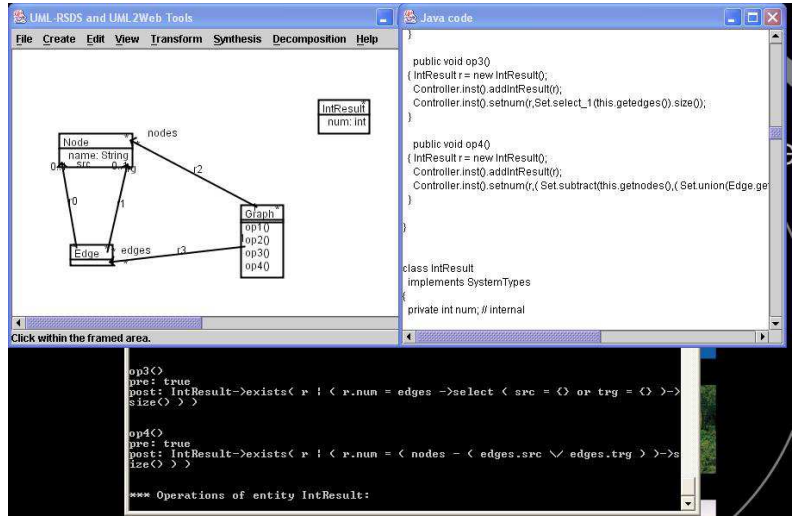}
\caption{Graph metamodel and queries in UML-RSDS}
\label{graphqmm}
\end{figure}

\begin{figure}[htbp]
\centering
\includegraphics[width=4.4in]{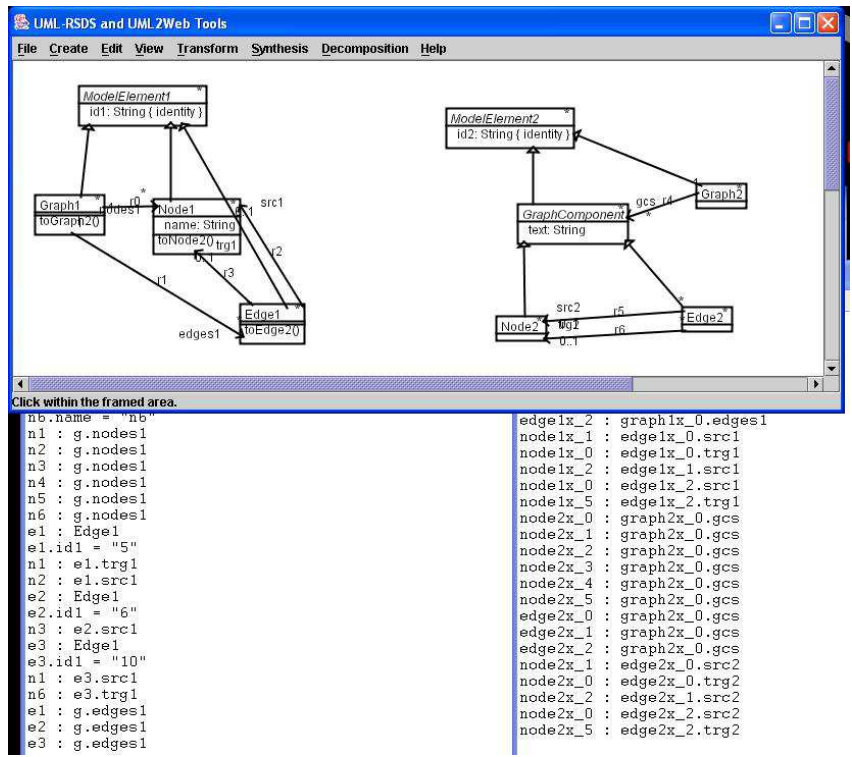}
\caption{Graph migration metamodels}
\label{smigmm}
\end{figure}

Source and target metamodels are defined using the
visual class diagram editor of UML-RSDS (Figures \ref{graphqmm} and \ref{smigmm}). Metamodels
cannot contain multiple inheritance, and all
non-leaf classes must be abstract. Metamodels can be
saved to a file by the $Save$ $data$ command.

Source models can be
defined in text files, which are then read by
the executable implementation of the transformation
metaclass, in a textual form. For example, a 
test model of the
simple graph metamodel can be defined as follows:
\begin{small}
\begin{verbatim}
g : Graph
n1 : Node 
n1.name = "n1"
n1 : g.nodes
n2 : Node
n2.name = "n2"
n2 : g.nodes
e : Edge
n1 : e.src
n2 : e.trg
e : g.edges
\end{verbatim}
\end{small}
This defines a single edge from the first to the second
node.
Alternative models can be defined in a similar way.

The UML-RSDS toolset is located at
{\tt http://www.dcs.kcl.ac.uk/staff/kcl/uml2web}.
UML-RSDS can be executed by the command 
{\tt java UmlTool}. The directory {\tt output} is used to
store metamodels, input and output
models, and the generated Java code. The command
$Load~data$ loads a metamodel from a file
(eg, $mig2.txt$ for the migration metamodel).
The command $Synthesis~Java$ generates the Java
executable of a transformation, this
generated executable is the 
{\tt Controller.java} file in the $output$ directory.
This can be compiled and used independently of the
toolset.


\small

\begin{thebibliography}{99}



 



\bibitem{helloworldcase} S. Mazanek,
{\em Hello World: an instructive case for the Transformation
Tool Contest}, in \cite{ttc2011eptcs}, 2011.




 


 


\bibitem{Lano10ifm} K. Lano, S. Kolahdouz-Rahimi,
{\em Specification and Verification of Model
Transformations using UML-RSDS},
IFM 2010.

\bibitem{mtpat}
K. Lano, S. Kolahdouz-Rahimi, 
{\em Model Transformation Design Patterns},
{ICSEA 2011.}

\bibitem{Lano11icmt} K. Lano, S. Kolahdouz-Rahimi,
{\em Model-Driven Development of Model
Transformations},
ICMT 2011.











\bibitem{ttc2011eptcs} Van Gorp, Pieter, Mazanek, Steffen,
and Rose, Louis,
{\em TTC 2011: Fifth Transformation Tool Contest,
Z\"urich, Switzerland, June 29-30 2011, Post-Proceedings},
EPTCS, 2011.

\end{thebibliography}
\end{document}